\documentclass[aps,prb,preprint,superscriptaddress]{revtex4}
\usepackage{amsmath}
\usepackage{amssymb}
\usepackage{amsthm}
\usepackage{amsfonts}
\usepackage{listings}
\usepackage{enumerate}
\usepackage{latexsym}
\usepackage{psfrag}
\usepackage{bm}
\usepackage[all]{xy}
\usepackage{graphicx}
\usepackage{subfigure}
\usepackage[pdftex,colorlinks=false]{hyperref}
\usepackage{color}
\usepackage{mathtools}
\usepackage{eufrak}
\usepackage{ gensymb }
\usepackage{lipsum}
\begin{document}

\title{Li doping kagome spin liquid compounds}
\author{Wei Jiang}
\affiliation{Department of Materials Science and Engineering, University of Utah, Salt Lake City, Utah 84112, USA}
\author{Huaqing Huang}
\affiliation{Department of Materials Science and Engineering, University of Utah, Salt Lake City, Utah 84112, USA}
\author{Jia-Wei Mei}
\email{meijw@sustc.edu.cn}
\affiliation{Institute for Quantum Science and Engineering, and Department of Physics, Southern University of Science and Technology, Shenzhen 518055, China}
\author{Feng Liu}
\email{fliu@eng.utah.edu}
\affiliation{Department of Materials Science and Engineering, University of Utah, Salt Lake City, Utah 84112, USA}

\date{\today}
\begin{abstract}
Herbertsmithite and Zn-doped barlowite are two compounds for experimental realization of two-dimensional gapped kagome spin liquid. Theoretically, it has been proposed that charge doping a quantum spin liquid gives rise to exotic metallic states, such as high-temperature superconductivity. However, one recent experiment about herbertsmithite with successful Li-doping shows surprisingly the insulating state even under the heavy doped scenario, which can hardly be explained by many-body physics. Using first-principles calculation, we performed a comprehensive study about the Li intercalated doping effect of these two compounds. For the Li-doped herbertsmithite, we identified the optimized Li position at the Cl-(OH)$_3$-Cl pentahedron site instead of previously speculated Cl-(OH)$_3$ tetrahedral site. With the increase of Li doping concentration, the saturation magnetization decreases linearly due to the charge transfer from Li to Cu ions. Moreover, we found that Li forms chemical bonds with the nearby (OH)$^-$ and Cl$^-$ ions, which lowers the surrounding chemical potential and traps the electron, as evidenced by the localized charge distribution, explaining the insulating behavior measured experimentally. Though with different structure from herbertsmithite, Zn-doped Barlowite shows the same features upon Li doping. We conclude that Li doping this family of kagome spin liquid cannot realize exotic metallic states, other methods should be further explored, such as element substitution with different valence electrons.
\end{abstract}

\maketitle

\section{Introduction}
When subject to strong geometric frustrations, quantum spin systems may achieve paramagnetic ground states dubbed resonance valance bond (RVB), or quantum spin liquid (QSL) states~\cite{Anderson1987}. QSL is an unambiguous Mott insulator whose charge gap is not associated with any symmetry breaking~\cite{Anderson1987} and it is characterized by the pattern of long-range quantum entanglement that has no classical counterpart~\cite{Wen2004,Kitaev2006,Levin2006}. Upon charge doping in the QSL, exotic quantum states may evolve, such as the high-temperature superconductivity as predicted in Anderson's doped RVB theory~\cite{Anderson1987}, metallic pseudogap state that has small hole-like Fermi pockets~\cite{Yang2006,Moon2011,Mei2012,Mei2012a} dubbed fractional Fermi liquid~\cite{Senthil2003,Senthil2004,Moon2011}, and Luttinger-volume violating Fermi liquid~\cite{Mei2012}.

Kagome Heisenberg antiferromagnets are promising systems for the pursuit of QSL~\cite{Lee2008,Balents2010,Norman2016,Depenbrock2012,Iqbal2013,Gong2015,Mei2017,He2017,Jiang2016,Liao2017}. Herbertsmithite (ZnCu$_3$(OH)$_6$Cl$_2$)~\cite{Shores2005,Helton2007,Mendels2007,Han2012,Fu2015} and Zn-doped barlowite (ZnCu$_3$(OH)$_6$FBr)~\cite{Feng2017a} are two reported compounds for two-dimensional (2D) realization of kagome spin liquid with gapped ground state. In herbertsmithite, inelastic neutron scattering measurements have detected continuum of spin excitations~\cite{Han2012} while nuclear magnetic resonance (NMR) measurements suggest a finite gap at low temperature~\cite{Fu2015}. On the other hand, NMR measurement about the Zn-doped barlowite~\cite{Feng2017a} demonstrates the gapped spin-1/2 spinon excitations.

Recently, Kelly \textit{et al.} have succeeded in the topochemical synthesis of Lithium intercalation doped herbertsmithite [ZnLi$_x$Cu$_3$(OH)$_6$Cl$_2$]. The electron from the intercalated Li is found indeed doped into the Cu$^{2+}$ kagome spin system as evidenced by the linear decrease of magnetization as a function of Li doping concentration. However, contrary to expectations, no metallicity or superconductivity was observed~\cite{Kelly2016}. It is natural to question whether the insulating behavior is due to the many-body physics or chemical reasons. From the many-body physics perspective, the insulating behavior in the lightly doped region can be explained using valance bond solid states~\cite{Guertler2011,Guertler2013} and holon Wigner crystal~\cite{Jiang2017} theory, which have been proposed based on the doped $t$-$J$ model for QSL. However, the insulating behavior remains even in the heavily doped region, e.g., $x=1.8$ with 3/5e per Cu$^{2+}$, which can be hardly explained by the many-body physics. For example, in Ref.~\onlinecite{Guertler2011}, a metallic state is expected with a large charge doping based on the variational Monte Carlo simulation. On the other hand, the chemical interactions in the Li-doped herbertsmithite has not been investigated yet, which is known to have critical influence to the inoic system and may provide an explanation to the insulating behavior with different doping concentration.

In this paper, using first-principles calculation method, we studied the Li intercalation doping effects of these two kagome spin liquid compounds. In the Li-doped herbertsmithite, it was speculated that Li is located in the Cl-(OH)$_3$ tetrahedron hole (T-site)~\cite{Kelly2016}, based on which singlet trapping and electron localization were proposed to explain the insulating behavior. However, we found that Li prefers to sit at the Cl-(OH)$_3$-Cl pentahedron site (square pyramid, P-site), having a total energy of around 0.8 eV per unit cell (u.c.) lower than that of the T-site. It is also found that the total magnetization decreases linearly with the increase of Li doping concentration, consistent with experiments, which is caused by the electron transfer from the intercalated Li to its nearby Cu ions. From the analysis of the projected density of states (PDOS), we found that Li forms bonding and antibonding states with the neighboring (OH)$^-$ and Cl$^-$ ions, which lowers the neighboring chemical potential and traps the doped electron in the vicinity of the doped Li, explaining the insulating behavior measured experimentally~\cite{Kelly2016}. Though the newly synthesized compound, Zn-doped barlowite, has different structure, we found that the doped Li still prefers to locate at the same location and forms chemical bond with nearby (OH)$^-$ and halogen ions, which is expected to result in the same insulating feature as Li-doped herbertsmithite.

\section{Calculation methods}
Our first-principles calculations were carried out within the framework of the Perdew-Burke-Ernzerhof generalized gradient approximation, as embedded in the Vienna $ab$ initio simulation package code~\cite{Kresse1993}. All the calculations were performed with a plane-wave cutoff energy of 500 eV. For structural relaxation, we adopted the experimental lattice constants for both herbertsmithite and Zn-doped barlowite, where the geometric optimizations were performed without any constraint until the force on each atom is less than 0.01eV\AA$^{-1}$ and the change of total energy is smaller than 10$^{-4}$ eV per unit cell. The Brillouin zone $k$-point sampling was set with a spacing of 0.03$\times$2$\pi$/\AA, which corresponds to a 6$\times$6$\times$4 and 3$\times$3$\times$4 $k$-point mesh for unit-cell and 2$\times$2$\times$1 supercell calculation, respectively. To better describe the localized 3$d$ electrons of Cu, an additional on-site Hubbard U term was added in the calculation of magnetic properties, with different U values tested.

\begin{figure}[tbp]
  \centering
  \includegraphics[width=\columnwidth]{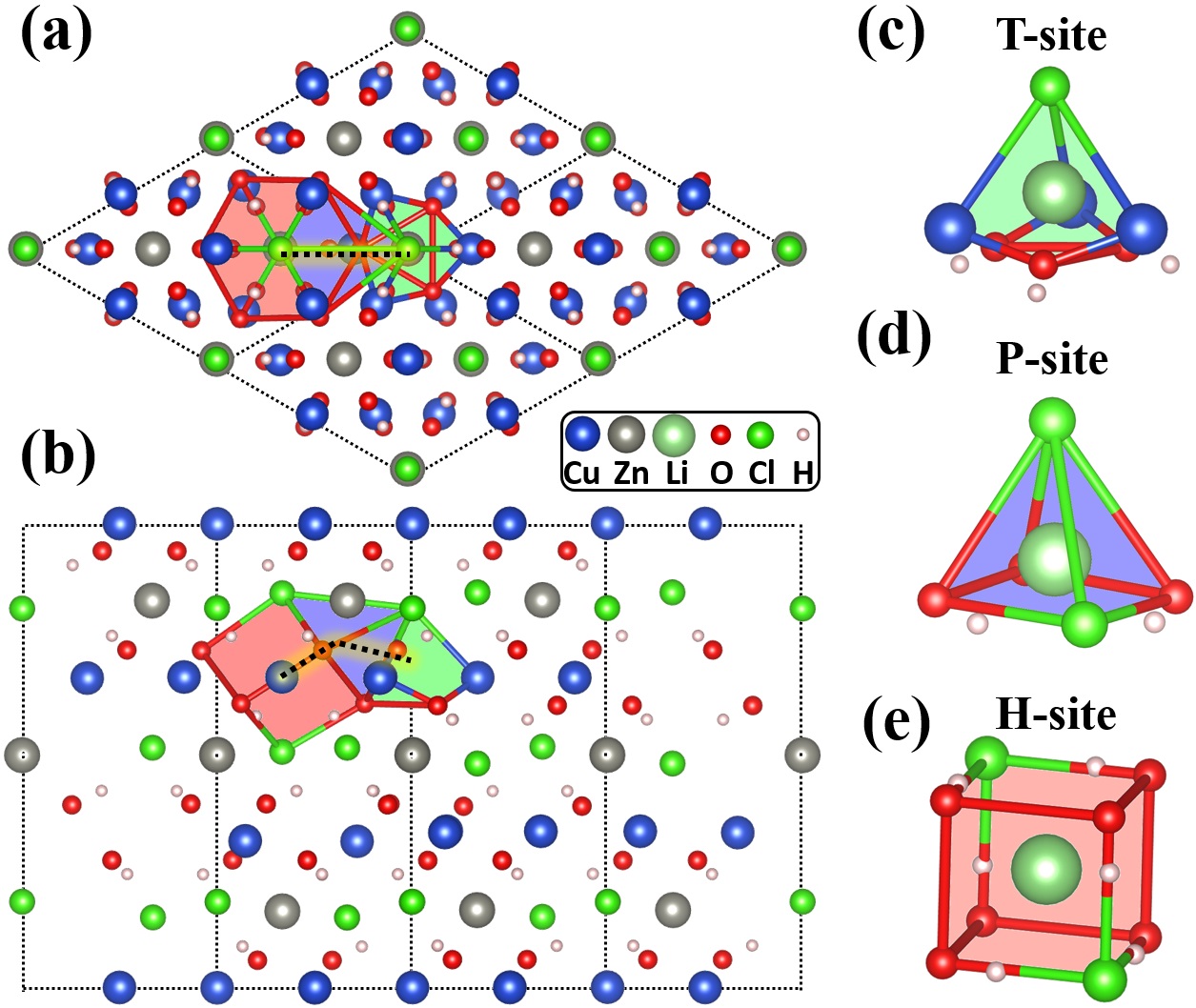}
  \caption{Li in herbertsmithite. (a) and (b) Top and side view of vacant space in the herbertsmithite with green, blue, and red colors highlighting the tetrahedron (T), pentahedron (P), and hexahedron (H) site, respectively. The dashed lines indicate the unit cell. (c), (d), and (e) Possible positions of Li ion in T-, P-, and H-site, respectively.}
  \label{fig:figure1}
\end{figure}

\section{Results and Discussion}

Herbertsmithite has the tetrahedral $R\bar{3}m$ (166) space group with $ABC$ stacked Cu$^{2+}$ spin-1/2 kagome planes along $c$ directions, which are separated by non-magnetic ions (Fig.~\ref{fig:figure1}). Different locations of possible Li occupancy have been considered in Ref.~\onlinecite{Kelly2016}, such as element replacement and interstitial insertion. The optimal site for Li ions in herbertsmithite was proposed to be the Cl-(OH)$_3$ T-site, as highlighted by the green polyhedron in Fig.~\ref{fig:figure1}. However, by analyzing the local atomic structure, the space of Cl-(OH)$_3$ T-site is too narrow for Li ions, which will cause a noticeable lattice distortion after Li insertion. There is more space in the nearby Cl-(OH)$_3$-Cl P-site and the Cl-(OH)$_4$-Cl hexahedron vacancy (H-site), as highlighted by blue and red polyhedron in Fig.~\ref{fig:figure1}, respectively. The pentahedron and the hexahedron share a common Cl-(OH)$_3$ square, while the tetrahedron and the pentahedron share a common Cl-O arris. Figure~\ref{fig:figure1}(c), (d), and (e) show the enlarged T-, P-, and H-sites for Li ions, respectively. It is straightforward to find that the concentration ratio for T-, P-, and H-site is 2:6:1, indicating the P-sites potentially have the largest capacity for doped Li ions.

\begin{figure}[b]
  \centering
  \includegraphics[width=\columnwidth]{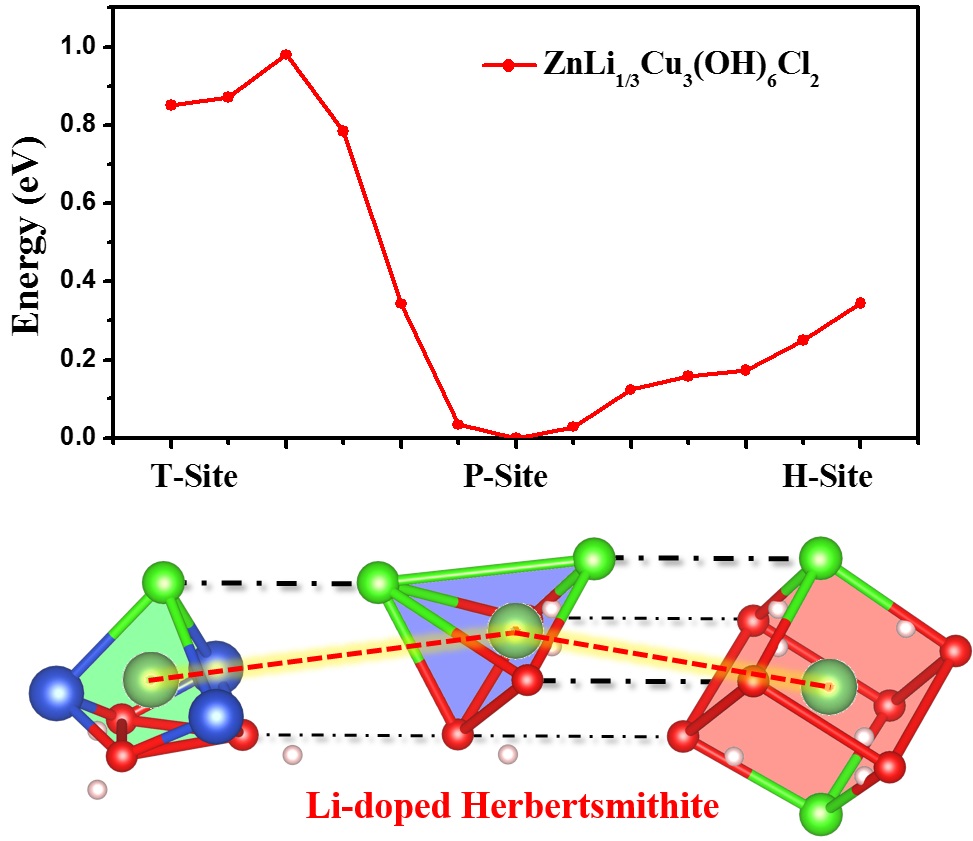}
  \caption{The nudged elastic band calculation for Li in herbertsmithite. The P-site is found to be the most stable position for Li. The highlighted red dashed lines indicate the path for the calculation.}
  \label{fig:figure2}
\end{figure}

To determine the most stable position for the doped Li ions in herbertsmithite, we first carried out total-energy calculation with Li ions in these three different locations, i.e. T-, P-, and H-sites. After relaxation, Li ion in the H-site is relaxed to the off-center position close to hydroxide (OH)$^-$, rather than the assumed cubic center, indicating Li prefers to bond with (OH)$^-$. We found the H-site with the largest space is not the most stable position for Li, instead, the position for Li with the lowest total energy is the P-site, formed by three (OH)$^-$ and two Cl ions, as shown in Fig.~\ref{fig:figure1}(d). The structure configuration with Li located in the T-site between Cl and the Cu plane is found metastable [Fig.~\ref{fig:figure1}(c)], which has around 0.8 eV/u.c. higher energy than that of T-site. It is important to mention that due to the small space of the T-site, the neighboring Cu and O atoms are greatly distorted after Li insertion, while atomic positions are nearly unchanged for the Li-doping in the P and H-site because of their larger space. To confirm the most stable position of Li in herbertsmithite, we then performed a nudged elastic band (NEB) calculation following the pathway from T- to P- to H-site, as shown by the highlighted black dash line in Fig.~\ref{fig:figure1}(a) and (b). The results are summarized in Fig.~\ref{fig:figure2}, showing that P-site is indeed the most stable position for Li with a total energy around 0.8 and 0.4 eV/u.c. lower than the T-site and H-site, respectively. We have further verified these results through 2$\times$2$\times$1 supercell calculations, which yield the same conclusion with the P-site to be the most stable position for Li.

\begin{figure}[t]
  \centering
   \includegraphics[width=\columnwidth]{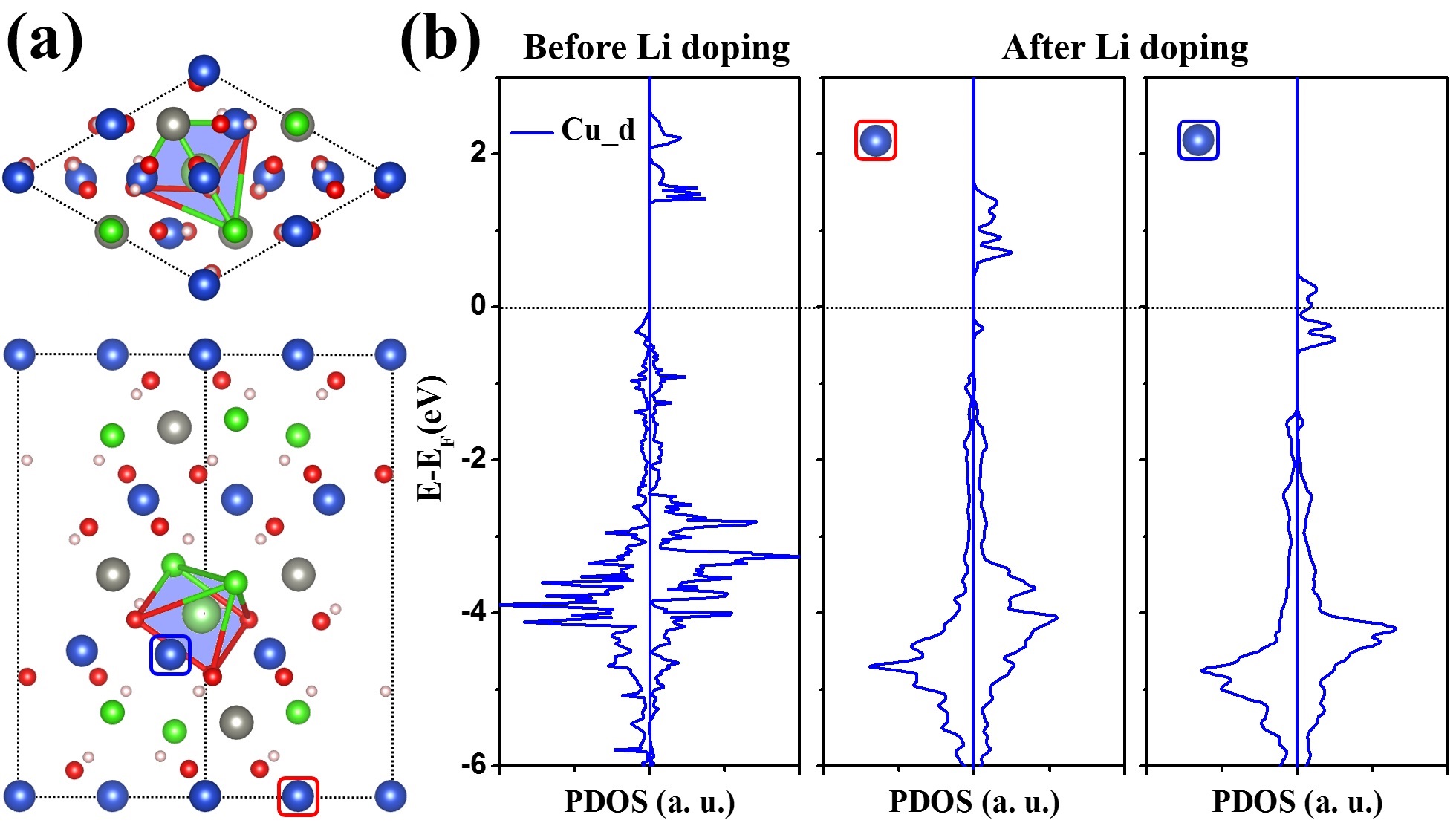}
  \caption{Li-doped herbertsmithite. (a) Top and side view of the most stable structure configuration for Li in herbertsmithite. The Cu ions in red and blue squares represent Cu ions that are unaffected and affected by the doped Li, respectively. (b) PDOS of Cu ions before and after the Li doping.}
  \label{fig:figure3}
\end{figure}

As described by the magnetic susceptibility measurements in the Ref.~\onlinecite{Kelly2016}, the magnetic moment of the Li-doped herbertsmithite decreases linearly with the increase of the Li doping concentration. Therefore, the electrons of Li are indeed doped into the Cu$^{2+}$ kagome planes. To demonstrate this behavior, we calculated the magnetic properties of the Li-doped herbertsmithite and plotted the PDOS of Cu before and after Li doping, using the most stable structural configuration for Li [Fig.~\ref{fig:figure3}(a)]. As shown in Fig.~\ref{fig:figure3}(b), before the doping of Li, the PDOS of Cu shows a gaped feature with the upper and lower Hubbard bands located above and below the Fermi level, respectively. After Li doping, the PDOS of the Cu ions away from the doped Li [Cu in red square in Fig.~\ref{fig:figure3}(a)] shows the similar gaped feature with moderate changes of the position of the Fermi level relative to the Cu DOS due to the doping effect. However, for the Cu close to the doped Li [Cu in blue square in Fig.~\ref{fig:figure3}(a)], the Fermi level crosses the upper Hubbard band of the PDOS, indicating the Cu gains electrons, which tune the Cu ions from Cu$^{2+}$ (S=1/2) into Cu$^{+}$ (S=0) to decrease the magnetic moment. Moreover, we studied the change of the total magnetic moment as a function of Li doping concentration by changing the number of Li ions in a 2$\times$2$\times$1 supercell. Using the ferromagnetic spin configuration to simulate the saturation magnetization, we observe the same linear decrease of magnetic moment as a function of Li doping concentration, as reported by the experiment magnetization measurement.

\begin{figure}[t]
  \centering
   \includegraphics[width=\columnwidth]{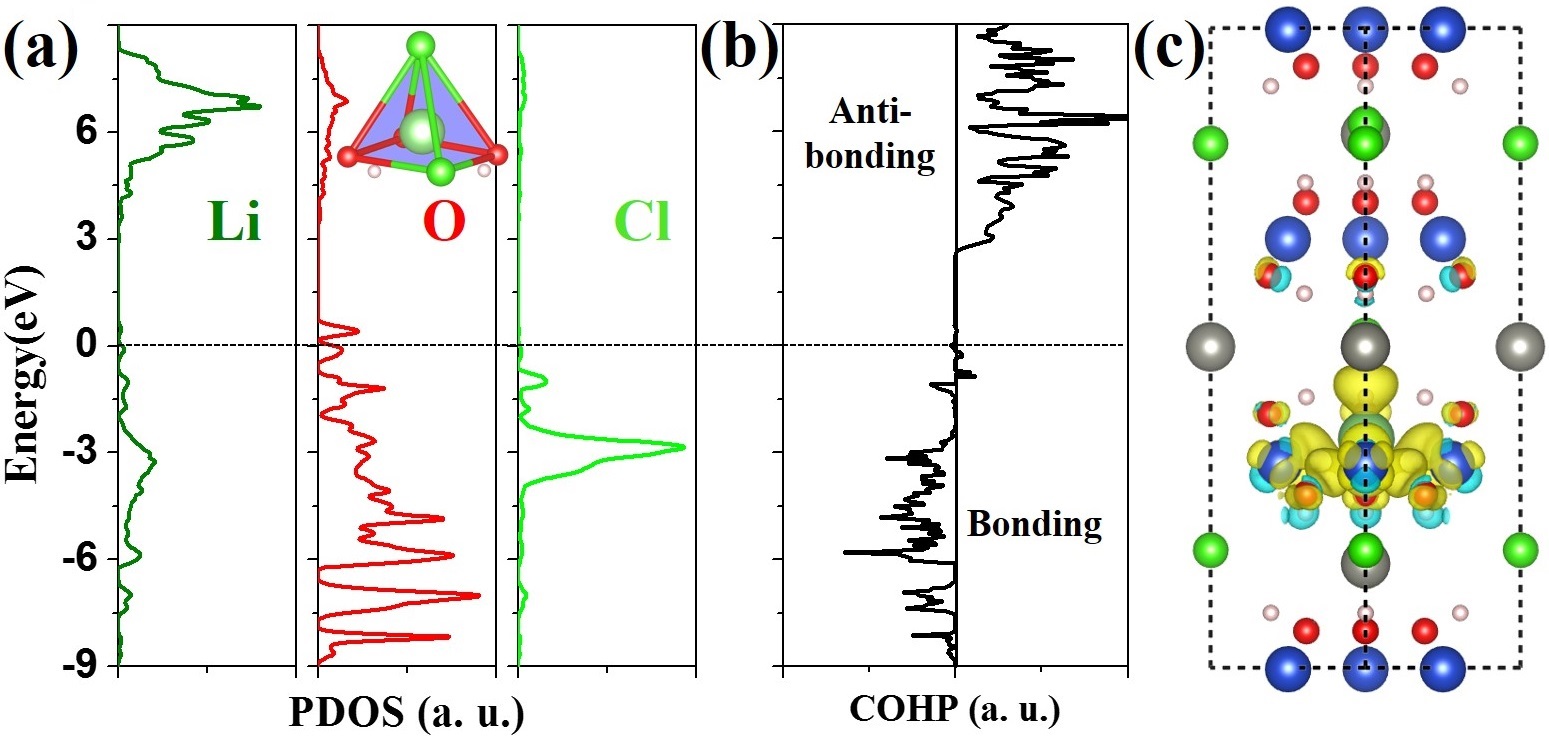}
  \caption{Chemical bonding in the Li-doped herbertsmithite. (a) Atomic-resolved PDOS. Inset shows the local environment of the Li in P-site. (b) COHP curve of the averaged Li-O and Li-Cl bonds. (c) Charge variance distribution.}
  \label{fig:figure4}
\end{figure}

The next mysterious question to be answered is how the doped electrons are trapped in the Cu kagome plane, which leads to the insulating feature observed experimentally. Several possible trapping mechanisms considering strong correlation have been proposed, such as singlet model and localized Cu triangle model based on the assumption that Li locates at the T-site~\cite{Kelly2016}. However, the optimized position for Li is actually in the P-site as discussed above and the many-body physics has not been found capable to explain the insulating behavior of the heavily doped kagome spin liquid~\cite{Guertler2011,Guertler2013,Jiang2017}. Therefore, there may exist chemical reasons for the electron trapping besides the many-body physics, such as chemical bonding between Li and Cl/OH$^-$ to form LiCl/Li(OH) pair to trap electrons, which can be properly captured by DFT. It is known that the chemical reactions in ionic systems are very sensitive to the location of the doped ions, therefore, we further studied detailed electronic properties of Li and its nearby OH$^-$ and Cl$^-$ ions with the most stable structure.

As shown in Fig.~\ref{fig:figure4}(a), we plotted the atomic-resolved PDOS of Li and its nearby Cl and O that form the pentagon cage to study their chemical reactions. From the PDOS of Li, it is clear that most of the DOS is located above the Fermi level (5-8 eV), confirming the electron transfer from Li ion to the nearby Cu ions. There are little DOS below the Fermi level with a relatively large broadening (-8-0 eV), showing the feature of strong hybridization. Interestingly, comparing the PDOS of O and Cl to that of Li, we find that their PDOS are located in the exactly same energy range, suggesting possible bondind and antibonding features. Different from the wide broadening of O $2p$ orbitals, the Cl $2p$ orbitals are more localized. Additionally, the chemical bonding between Li-O and Li-Cl were investigated using the crystal orbital Hamilton population (COHP) analysis~\cite{Dronskowski1993}, as shown in the Fig.~\ref{fig:figure4}(b). As observed from the averaged COHP curve, the bonding and anti-bonding feature is prominent for states below and above the Fermi level, respectively. These chemical reactions lead to lower chemical potential in the vicinity of the doped Li that traps the electrons on the nearby Cu ions, yielding the insulating feature.

\begin{figure}[tbp]
  \centering
   \includegraphics[width=\columnwidth]{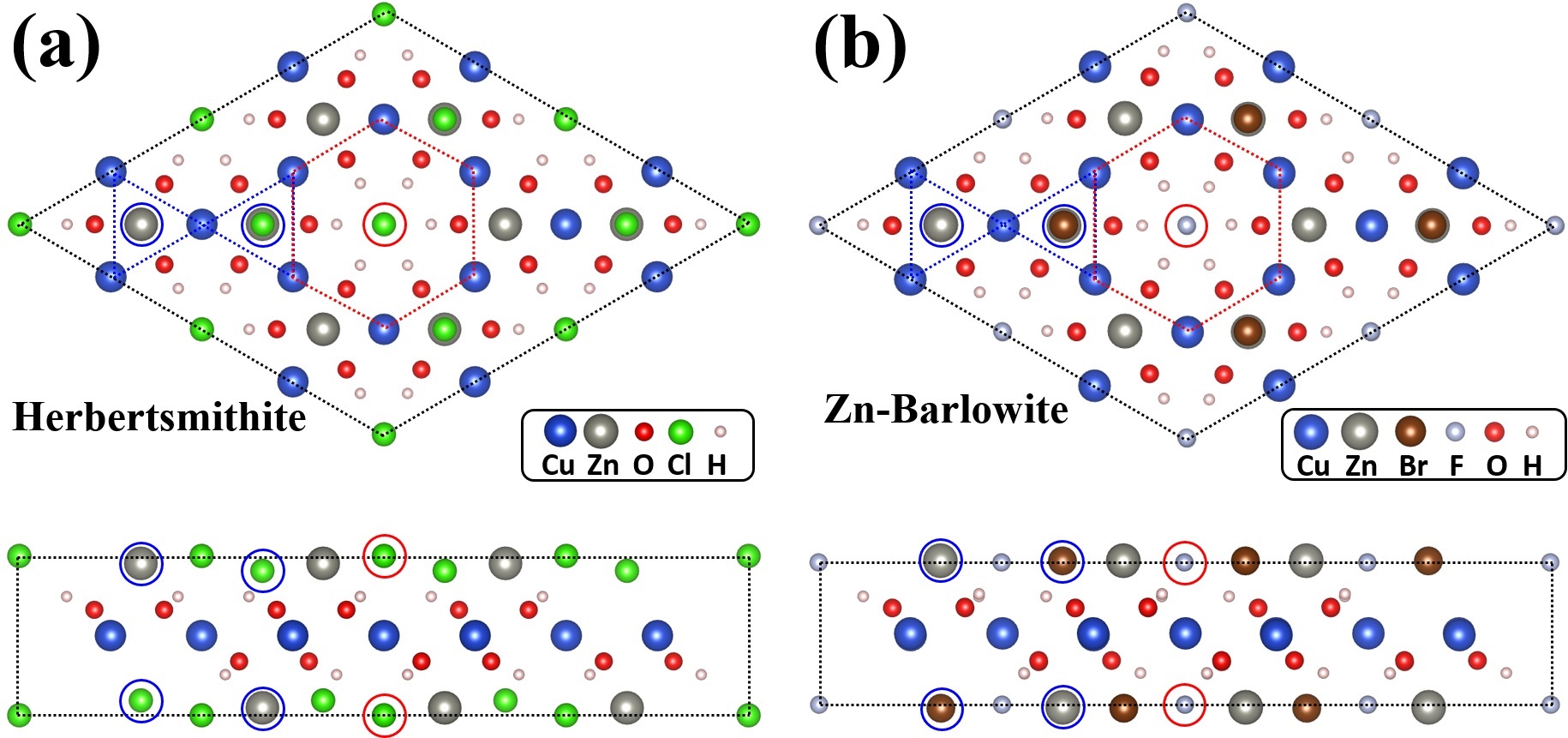}
  \caption{Local environment of the Cu kagome plane in herbertsmithite and Zn-doped barlowite. (a) and (b) Top and side view of the Cu kagome plane in herbertsmithite and Zn-doped barlowite, respectively. For herbertsmithite/Zn-doped barlowite, halogen atoms Cl/F (red circle) sit on top of the Cu hexagon center (red hexagon) and the space above and below the Cu triangle (blue triangle) is occupied by Zn and Cl/Br (blue circles) alternatively.}
  \label{fig:figure5}
\end{figure}

To further support our conjecture, we plotted the charge variance distribution caused by the Li doping. As shown in Fig.~\ref{fig:figure4}(c), it can be clearly seen that the doped electron is mainly distributed around the Li ion and its neighboring ions, confirming the electron localization caused by chemical bonding around the Li ion. We additionally examined the size effect using a larger 2$\times$2$\times$1 supercell, and found similar feature of electron localization. It is worth noting that because of the large capacity of the P-site to hold Li ions with relatively small structure distortion, the structure remains stable in a wide range of Li doping concentration. Unlike other proposals, where the electron doped kagome spin liquid shows distinct behaviors under different doping concentrations, i.e. insulating and metallic state under lightly and heavily doped condition, respectively, the chemical reasons may affect the system consistently in a wide range of Li doping concentration before the structural deformation. We believe this could explain why the system was found remaining insulating experimentally under different Li doping concentrations.

\begin{figure}[tbp]
  \centering
   \includegraphics[width=\columnwidth]{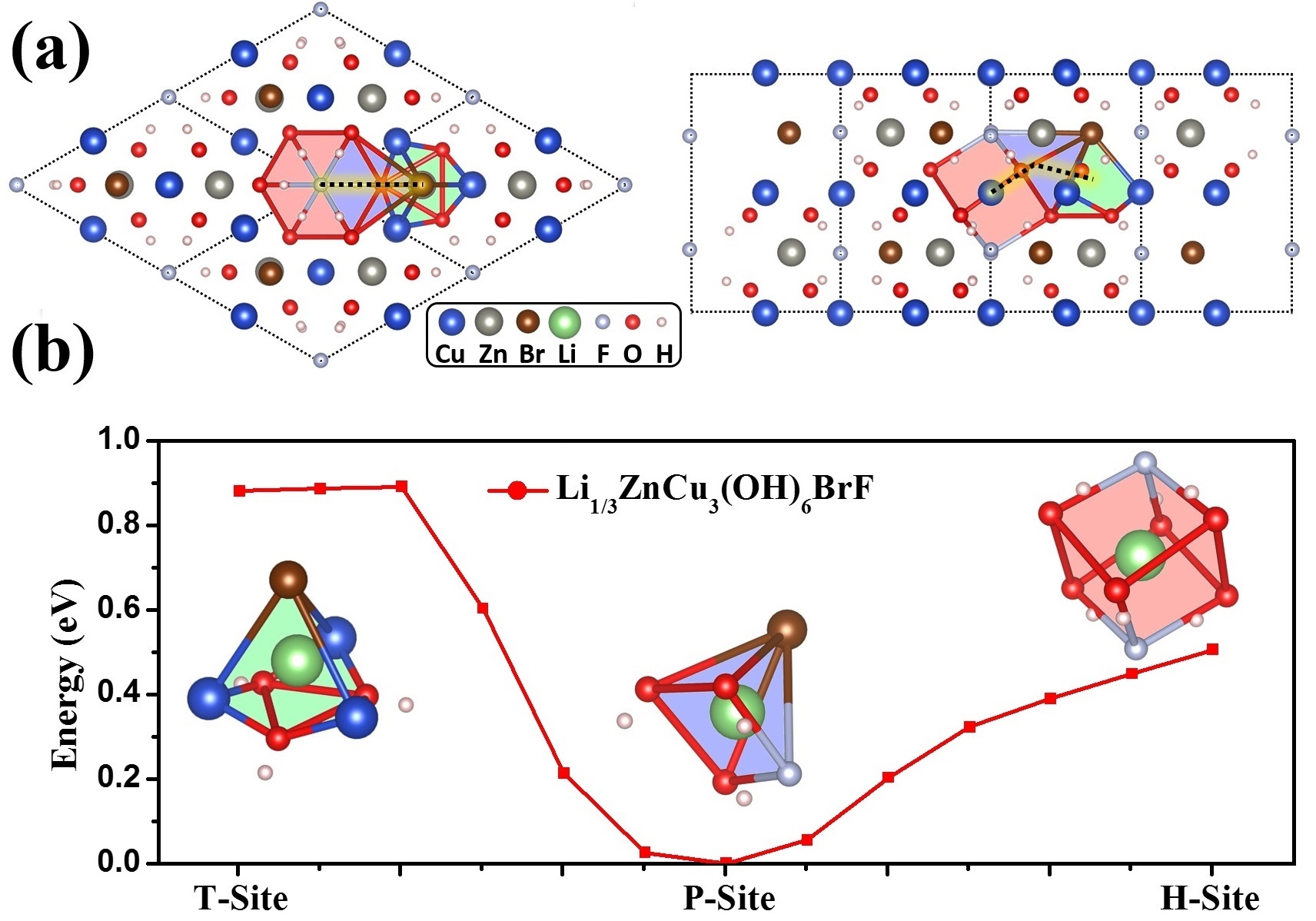}
  \caption{Li positions in Zn-doped barlowite. (a) and (b) Top and side view of vacant spaces in the Zn-doped barlowite with green, blue, and red colors highlighting the T-, P-, and H-site, respectively. (c) The NEB calculation results for possible positions of Li ion along the path from T-site to P-site to H-site. }
  \label{fig:figure6}
\end{figure}

Very recently, Zn-doped barlowite has been successfully synthesized and a gaped quantum spin liquid ground state is revealed in Ref.~\onlinecite{Feng2017a}. Distinguished from the herbertsmithite with $ABC$-stacked kagome planes, Zn-doped barlowite has $AA$-stacked Cu$^{2+}$ kagome planes along $c$ direction with the hexagonal $P6_3/mmc$ (193) space group. Nevertheless, we found that the kagome planes formed by Cu and O are identical in these two materials with very similar local environments. As can be seen from Fig.~\ref{fig:figure5}, for both materials, halogen atoms sit on the top of Cu hexagon center, and the space above the Cu triangle center is occupied by Zn and halogen atoms alternatively. Moreover, T-, P-, and H-sites in the Zn-doped barlowite are next to each other with a site-ratio of 2:6:1 [Fig.~\ref{fig:figure6}(a)], which are the same as that of the herbertsmithite. Differently, halogen atom Cl in the herbertsmithite is replaced by Br and F ions in the Zn-doped barlowite, as shown in Fig.~\ref{fig:figure5}. Based on the total energy calculations for Li in different vacancies of the Zn-doped barlowite, we found the P-site is also the most stable location for Li ions, having around 0.9 and 0.5 eV/u.c. lower energy than the T- and H-site, respectively. The NEB calculation results are summarized in Fig.~\ref{fig:figure6}(b), further confirming the most stable site of P-site and the metastable state of T-site for Li in the Zn-doped barlowite.

We then calculated the magnetic properties of the Zn-doped barlowite with different Li doping concentrations. The magnetic behavior of the Zn-doped barlowite upon Li doping is the same as that of the herbertsmithite, where a linear decrease of magnetization is observed with the increase of Li doping concentration, caused by the charge transfer from Li ions to its nearby Cu ions. Similarly, through PDOS and COHP analysis, we found that Li forms chemical bonds with the neighboring OH$^-$, Br$^-$, and F$^-$ ions, which lowers the chemical potential of its nearby region, leading to the electron localization. Therefore, we expect that for the Zn-doped barlowite, the Li doping will also yield the same insulating behavior as that of the Li-doped herbertsmithite, even in the heavily doped region. For the same reason, other interstitial doping methods with different elements would not succeed either. In order to realize exotic metallic states, i.e. dope electrons into the Cu kagome plane without any restriction, the doping method must avoid chemical changes close to the Cu kagome plane or involve chemical changes that have negligible effect to the Cu ions. To meet this requirement, substitution of the inter-kagome plane metal ion, Zinc, using other metal ions with similar radius but different electronic configurations may be feasible.

\section{Conclusions}
In conclusion, we have carried out a comprehensive study about the Li intercalated doping effect of the two promising compounds for 2D realization of kagome spin liquid, i.e. herbertsmishite and Zn-doped barlowite. Though with different layer stacking sequence for the Cu kagome planes, the interstitial vacancies are surprisingly alike for these two compounds, which leads to very similar behaviors upon Li doping. Instead of the previous speculated T-site, P-site was found to be the most stable position for the doped Li ions, which has around 0.8 and 0.9 eV/u.c. lower energy than Li in the T-site for herbertsmithite and Zn-doped barlowite, respectively. We demonstrated the charge transfer from the doped Li to its adjacent Cu ions, which changes Cu$^{2+}$ (S=1/2) into Cu$^+$ (S=0) and leads to the linear decrease of the saturation magnetization as a function of Li doping concentration. Due to the chemical bonding formed between Li and its nearby OH$^-$, Cl$^-$, and F$^-$ ions, the chemical potential in the vicinity of Li is lowered, which affects the nearby Cu ions and causes electron localization, giving rise to the insulating behavior even under heavily doped region. Like many other normal ionic compounds, whose electronic and magnetic properties are greatly affected by chemical reactions upon ionic doping, we have demonstrated that chemical reactions also play a critical role in determining the behaviors of this family of kagome spin liquid compounds with strong correlation. Finally, we propose that to realize exotic metallic state, suitable element substitution that avoid chemical changes near the Cu kagome plane may be promising. Our study paves the way for studies of electron doping a kagome spin liquid.

\section{Acknowledgements}
H. Huang and F. Liu are supported by U.S. DOE-BES (Grant No. DE-FG02-04ER46148). W. Jiang is supported by the National Science Foundation-Material Research Science \& Engineering Center (NSF-MRSEC grand No. DMR-1121252). We also thank the CHPC at the University of Utah and DOE-NERSC for providing the computing resources.

\bibliography{LiQSL}

\begin{thebibliography}{32}
\expandafter\ifx\csname natexlab\endcsname\relax\def\natexlab#1{#1}\fi
\expandafter\ifx\csname bibnamefont\endcsname\relax
  \def\bibnamefont#1{#1}\fi
\expandafter\ifx\csname bibfnamefont\endcsname\relax
  \def\bibfnamefont#1{#1}\fi
\expandafter\ifx\csname citenamefont\endcsname\relax
  \def\citenamefont#1{#1}\fi
\expandafter\ifx\csname url\endcsname\relax
  \def\url#1{\texttt{#1}}\fi
\expandafter\ifx\csname urlprefix\endcsname\relax\def\urlprefix{URL }\fi
\providecommand{\bibinfo}[2]{#2}
\providecommand{\eprint}[2][]{\url{#2}}

\bibitem[{\citenamefont{Anderson}(1987)}]{Anderson1987}
\bibinfo{author}{\bibfnamefont{P.~W.} \bibnamefont{Anderson}},
  \bibinfo{journal}{Science} \textbf{\bibinfo{volume}{235}},
  \bibinfo{pages}{1196} (\bibinfo{year}{1987}).

\bibitem[{\citenamefont{Wen}(2004)}]{Wen2004}
\bibinfo{author}{\bibfnamefont{X.}~\bibnamefont{Wen}},
  \emph{\bibinfo{title}{{Quantum Field Theory of Many-Body Systems:From the
  Origin of Sound to an Origin of Light and Electrons}}}, Oxford Graduate Texts
  (\bibinfo{publisher}{OUP Oxford}, \bibinfo{year}{2004}), ISBN
  \bibinfo{isbn}{9780198530947}.

\bibitem[{\citenamefont{Kitaev and Preskill}(2006)}]{Kitaev2006}
\bibinfo{author}{\bibfnamefont{A.}~\bibnamefont{Kitaev}} \bibnamefont{and}
  \bibinfo{author}{\bibfnamefont{J.}~\bibnamefont{Preskill}},
  \bibinfo{journal}{Phys. Rev. Lett.} \textbf{\bibinfo{volume}{96}},
  \bibinfo{pages}{110404} (\bibinfo{year}{2006}).

\bibitem[{\citenamefont{Levin and Wen}(2006)}]{Levin2006}
\bibinfo{author}{\bibfnamefont{M.}~\bibnamefont{Levin}} \bibnamefont{and}
  \bibinfo{author}{\bibfnamefont{X.-G.} \bibnamefont{Wen}},
  \bibinfo{journal}{Phys. Rev. Lett.} \textbf{\bibinfo{volume}{96}},
  \bibinfo{pages}{110405} (\bibinfo{year}{2006}).

\bibitem[{\citenamefont{Yang et~al.}(2006)\citenamefont{Yang, Rice, and
  Zhang}}]{Yang2006}
\bibinfo{author}{\bibfnamefont{K.-Y.} \bibnamefont{Yang}},
  \bibinfo{author}{\bibfnamefont{T.~M.} \bibnamefont{Rice}}, \bibnamefont{and}
  \bibinfo{author}{\bibfnamefont{F.-C.} \bibnamefont{Zhang}},
  \bibinfo{journal}{Phys. Rev. B} \textbf{\bibinfo{volume}{73}},
  \bibinfo{pages}{174501} (\bibinfo{year}{2006}),
  \urlprefix\url{https://link.aps.org/doi/10.1103/PhysRevB.73.174501}.

\bibitem[{\citenamefont{Moon and Sachdev}(2011)}]{Moon2011}
\bibinfo{author}{\bibfnamefont{E.~G.} \bibnamefont{Moon}} \bibnamefont{and}
  \bibinfo{author}{\bibfnamefont{S.}~\bibnamefont{Sachdev}},
  \bibinfo{journal}{Phys. Rev. B} \textbf{\bibinfo{volume}{83}},
  \bibinfo{pages}{224508} (\bibinfo{year}{2011}),
  \urlprefix\url{https://link.aps.org/doi/10.1103/PhysRevB.83.224508}.

\bibitem[{\citenamefont{Mei et~al.}(2012)\citenamefont{Mei, Kawasaki, Zheng,
  Weng, and Wen}}]{Mei2012}
\bibinfo{author}{\bibfnamefont{J.-W.} \bibnamefont{Mei}},
  \bibinfo{author}{\bibfnamefont{S.}~\bibnamefont{Kawasaki}},
  \bibinfo{author}{\bibfnamefont{G.-Q.} \bibnamefont{Zheng}},
  \bibinfo{author}{\bibfnamefont{Z.-Y.} \bibnamefont{Weng}}, \bibnamefont{and}
  \bibinfo{author}{\bibfnamefont{X.-G.} \bibnamefont{Wen}},
  \bibinfo{journal}{Phys. Rev. B} \textbf{\bibinfo{volume}{85}},
  \bibinfo{pages}{134519} (\bibinfo{year}{2012}),
  \urlprefix\url{https://link.aps.org/doi/10.1103/PhysRevB.85.134519}.

\bibitem[{\citenamefont{Mei}(2012)}]{Mei2012a}
\bibinfo{author}{\bibfnamefont{J.-W.} \bibnamefont{Mei}},
  \bibinfo{journal}{Phys. Rev. Lett.} \textbf{\bibinfo{volume}{108}},
  \bibinfo{pages}{227207} (\bibinfo{year}{2012}),
  \urlprefix\url{https://link.aps.org/doi/10.1103/PhysRevLett.108.227207}.

\bibitem[{\citenamefont{Senthil et~al.}(2003)\citenamefont{Senthil, Sachdev,
  and Vojta}}]{Senthil2003}
\bibinfo{author}{\bibfnamefont{T.}~\bibnamefont{Senthil}},
  \bibinfo{author}{\bibfnamefont{S.}~\bibnamefont{Sachdev}}, \bibnamefont{and}
  \bibinfo{author}{\bibfnamefont{M.}~\bibnamefont{Vojta}},
  \bibinfo{journal}{Phys. Rev. Lett.} \textbf{\bibinfo{volume}{90}},
  \bibinfo{pages}{216403} (\bibinfo{year}{2003}),
  \urlprefix\url{https://link.aps.org/doi/10.1103/PhysRevLett.90.216403}.

\bibitem[{\citenamefont{Senthil et~al.}(2004)\citenamefont{Senthil, Vojta, and
  Sachdev}}]{Senthil2004}
\bibinfo{author}{\bibfnamefont{T.}~\bibnamefont{Senthil}},
  \bibinfo{author}{\bibfnamefont{M.}~\bibnamefont{Vojta}}, \bibnamefont{and}
  \bibinfo{author}{\bibfnamefont{S.}~\bibnamefont{Sachdev}},
  \bibinfo{journal}{Phys. Rev. B} \textbf{\bibinfo{volume}{69}},
  \bibinfo{pages}{035111} (\bibinfo{year}{2004}),
  \urlprefix\url{https://link.aps.org/doi/10.1103/PhysRevB.69.035111}.

\bibitem[{\citenamefont{Lee}(2008)}]{Lee2008}
\bibinfo{author}{\bibfnamefont{P.~A.} \bibnamefont{Lee}},
  \bibinfo{journal}{Science} \textbf{\bibinfo{volume}{321}},
  \bibinfo{pages}{1306} (\bibinfo{year}{2008}).

\bibitem[{\citenamefont{Balents}(2010)}]{Balents2010}
\bibinfo{author}{\bibfnamefont{L.}~\bibnamefont{Balents}},
  \bibinfo{journal}{Nature} \textbf{\bibinfo{volume}{464}},
  \bibinfo{pages}{199} (\bibinfo{year}{2010}).

\bibitem[{\citenamefont{Norman}(2016)}]{Norman2016}
\bibinfo{author}{\bibfnamefont{M.~R.} \bibnamefont{Norman}},
  \bibinfo{journal}{Rev. Mod. Phys.} \textbf{\bibinfo{volume}{88}},
  \bibinfo{pages}{041002} (\bibinfo{year}{2016}).

\bibitem[{\citenamefont{Depenbrock et~al.}(2012)\citenamefont{Depenbrock,
  McCulloch, and Schollw\"ock}}]{Depenbrock2012}
\bibinfo{author}{\bibfnamefont{S.}~\bibnamefont{Depenbrock}},
  \bibinfo{author}{\bibfnamefont{I.~P.} \bibnamefont{McCulloch}},
  \bibnamefont{and}
  \bibinfo{author}{\bibfnamefont{U.}~\bibnamefont{Schollw\"ock}},
  \bibinfo{journal}{Phys. Rev. Lett.} \textbf{\bibinfo{volume}{109}},
  \bibinfo{pages}{067201} (\bibinfo{year}{2012}),
  \urlprefix\url{http://link.aps.org/doi/10.1103/PhysRevLett.109.067201}.

\bibitem[{\citenamefont{Iqbal et~al.}(2013)\citenamefont{Iqbal, Becca, Sorella,
  and Poilblanc}}]{Iqbal2013}
\bibinfo{author}{\bibfnamefont{Y.}~\bibnamefont{Iqbal}},
  \bibinfo{author}{\bibfnamefont{F.}~\bibnamefont{Becca}},
  \bibinfo{author}{\bibfnamefont{S.}~\bibnamefont{Sorella}}, \bibnamefont{and}
  \bibinfo{author}{\bibfnamefont{D.}~\bibnamefont{Poilblanc}},
  \bibinfo{journal}{Phys. Rev. B} \textbf{\bibinfo{volume}{87}},
  \bibinfo{pages}{060405} (\bibinfo{year}{2013}),
  \urlprefix\url{https://link.aps.org/doi/10.1103/PhysRevB.87.060405}.

\bibitem[{\citenamefont{Gong et~al.}(2015)\citenamefont{Gong, Zhu, Balents, and
  Sheng}}]{Gong2015}
\bibinfo{author}{\bibfnamefont{S.-S.} \bibnamefont{Gong}},
  \bibinfo{author}{\bibfnamefont{W.}~\bibnamefont{Zhu}},
  \bibinfo{author}{\bibfnamefont{L.}~\bibnamefont{Balents}}, \bibnamefont{and}
  \bibinfo{author}{\bibfnamefont{D.~N.} \bibnamefont{Sheng}},
  \bibinfo{journal}{Phys. Rev. B} \textbf{\bibinfo{volume}{91}},
  \bibinfo{pages}{075112} (\bibinfo{year}{2015}),
  \urlprefix\url{https://link.aps.org/doi/10.1103/PhysRevB.91.075112}.

\bibitem[{\citenamefont{Mei et~al.}(2017)\citenamefont{Mei, Chen, He, and
  Wen}}]{Mei2017}
\bibinfo{author}{\bibfnamefont{J.-W.} \bibnamefont{Mei}},
  \bibinfo{author}{\bibfnamefont{J.-Y.} \bibnamefont{Chen}},
  \bibinfo{author}{\bibfnamefont{H.}~\bibnamefont{He}}, \bibnamefont{and}
  \bibinfo{author}{\bibfnamefont{X.-G.} \bibnamefont{Wen}},
  \bibinfo{journal}{Phys. Rev. B} \textbf{\bibinfo{volume}{95}},
  \bibinfo{pages}{235107} (\bibinfo{year}{2017}),
  \urlprefix\url{https://link.aps.org/doi/10.1103/PhysRevB.95.235107}.

\bibitem[{\citenamefont{He et~al.}(2017)\citenamefont{He, Zaletel, Oshikawa,
  and Pollmann}}]{He2017}
\bibinfo{author}{\bibfnamefont{Y.-C.} \bibnamefont{He}},
  \bibinfo{author}{\bibfnamefont{M.~P.} \bibnamefont{Zaletel}},
  \bibinfo{author}{\bibfnamefont{M.}~\bibnamefont{Oshikawa}}, \bibnamefont{and}
  \bibinfo{author}{\bibfnamefont{F.}~\bibnamefont{Pollmann}},
  \bibinfo{journal}{Phys. Rev. X} \textbf{\bibinfo{volume}{7}},
  \bibinfo{pages}{031020} (\bibinfo{year}{2017}),
  \urlprefix\url{https://link.aps.org/doi/10.1103/PhysRevX.7.031020}.

\bibitem[{\citenamefont{Jiang et~al.}(2016)\citenamefont{Jiang, Kim, Han, and
  Ran}}]{Jiang2016}
\bibinfo{author}{\bibfnamefont{S.}~\bibnamefont{Jiang}},
  \bibinfo{author}{\bibfnamefont{P.}~\bibnamefont{Kim}},
  \bibinfo{author}{\bibfnamefont{J.~H.} \bibnamefont{Han}}, \bibnamefont{and}
  \bibinfo{author}{\bibfnamefont{Y.}~\bibnamefont{Ran}},
  \bibinfo{journal}{arXiv preprint arXiv:1610.02024}  (\bibinfo{year}{2016}).

\bibitem[{\citenamefont{Liao et~al.}(2017)\citenamefont{Liao, Xie, Chen, Liu,
  Xie, Huang, Normand, and Xiang}}]{Liao2017}
\bibinfo{author}{\bibfnamefont{H.~J.} \bibnamefont{Liao}},
  \bibinfo{author}{\bibfnamefont{Z.~Y.} \bibnamefont{Xie}},
  \bibinfo{author}{\bibfnamefont{J.}~\bibnamefont{Chen}},
  \bibinfo{author}{\bibfnamefont{Z.~Y.} \bibnamefont{Liu}},
  \bibinfo{author}{\bibfnamefont{H.~D.} \bibnamefont{Xie}},
  \bibinfo{author}{\bibfnamefont{R.~Z.} \bibnamefont{Huang}},
  \bibinfo{author}{\bibfnamefont{B.}~\bibnamefont{Normand}}, \bibnamefont{and}
  \bibinfo{author}{\bibfnamefont{T.}~\bibnamefont{Xiang}},
  \bibinfo{journal}{Phys. Rev. Lett.} \textbf{\bibinfo{volume}{118}},
  \bibinfo{pages}{137202} (\bibinfo{year}{2017}),
  \urlprefix\url{https://link.aps.org/doi/10.1103/PhysRevLett.118.137202}.

\bibitem[{\citenamefont{Shores et~al.}(2005)\citenamefont{Shores, Nytko,
  Bartlett, and Nocera}}]{Shores2005}
\bibinfo{author}{\bibfnamefont{M.~P.} \bibnamefont{Shores}},
  \bibinfo{author}{\bibfnamefont{E.~A.} \bibnamefont{Nytko}},
  \bibinfo{author}{\bibfnamefont{B.~M.} \bibnamefont{Bartlett}},
  \bibnamefont{and} \bibinfo{author}{\bibfnamefont{D.~G.}
  \bibnamefont{Nocera}}, \bibinfo{journal}{J. Am. Chem. Soc.}
  \textbf{\bibinfo{volume}{127}}, \bibinfo{pages}{13462}
  (\bibinfo{year}{2005}).

\bibitem[{\citenamefont{Helton et~al.}(2007)\citenamefont{Helton, Matan,
  Shores, Nytko, Bartlett, Yoshida, Takano, Suslov, Qiu, Chung
  et~al.}}]{Helton2007}
\bibinfo{author}{\bibfnamefont{J.~S.} \bibnamefont{Helton}},
  \bibinfo{author}{\bibfnamefont{K.}~\bibnamefont{Matan}},
  \bibinfo{author}{\bibfnamefont{M.~P.} \bibnamefont{Shores}},
  \bibinfo{author}{\bibfnamefont{E.~A.} \bibnamefont{Nytko}},
  \bibinfo{author}{\bibfnamefont{B.~M.} \bibnamefont{Bartlett}},
  \bibinfo{author}{\bibfnamefont{Y.}~\bibnamefont{Yoshida}},
  \bibinfo{author}{\bibfnamefont{Y.}~\bibnamefont{Takano}},
  \bibinfo{author}{\bibfnamefont{A.}~\bibnamefont{Suslov}},
  \bibinfo{author}{\bibfnamefont{Y.}~\bibnamefont{Qiu}},
  \bibinfo{author}{\bibfnamefont{J.-H.} \bibnamefont{Chung}},
  \bibnamefont{et~al.}, \bibinfo{journal}{Phys. Rev. Lett.}
  \textbf{\bibinfo{volume}{98}}, \bibinfo{pages}{107204}
  (\bibinfo{year}{2007}).

\bibitem[{\citenamefont{Mendels et~al.}(2007)\citenamefont{Mendels, Bert,
  de~Vries, Olariu, Harrison, Duc, Trombe, Lord, Amato, and
  Baines}}]{Mendels2007}
\bibinfo{author}{\bibfnamefont{P.}~\bibnamefont{Mendels}},
  \bibinfo{author}{\bibfnamefont{F.}~\bibnamefont{Bert}},
  \bibinfo{author}{\bibfnamefont{M.~A.} \bibnamefont{de~Vries}},
  \bibinfo{author}{\bibfnamefont{A.}~\bibnamefont{Olariu}},
  \bibinfo{author}{\bibfnamefont{A.}~\bibnamefont{Harrison}},
  \bibinfo{author}{\bibfnamefont{F.}~\bibnamefont{Duc}},
  \bibinfo{author}{\bibfnamefont{J.~C.} \bibnamefont{Trombe}},
  \bibinfo{author}{\bibfnamefont{J.~S.} \bibnamefont{Lord}},
  \bibinfo{author}{\bibfnamefont{A.}~\bibnamefont{Amato}}, \bibnamefont{and}
  \bibinfo{author}{\bibfnamefont{C.}~\bibnamefont{Baines}},
  \bibinfo{journal}{Phys. Rev. Lett.} \textbf{\bibinfo{volume}{98}},
  \bibinfo{pages}{077204} (\bibinfo{year}{2007}).

\bibitem[{\citenamefont{Han et~al.}(2012)\citenamefont{Han, Helton, Chu,
  Nocera, Rodriguez-Rivera, Broholm, and Lee}}]{Han2012}
\bibinfo{author}{\bibfnamefont{T.-H.} \bibnamefont{Han}},
  \bibinfo{author}{\bibfnamefont{J.~S.} \bibnamefont{Helton}},
  \bibinfo{author}{\bibfnamefont{S.}~\bibnamefont{Chu}},
  \bibinfo{author}{\bibfnamefont{D.~G.} \bibnamefont{Nocera}},
  \bibinfo{author}{\bibfnamefont{J.~A.} \bibnamefont{Rodriguez-Rivera}},
  \bibinfo{author}{\bibfnamefont{C.}~\bibnamefont{Broholm}}, \bibnamefont{and}
  \bibinfo{author}{\bibfnamefont{Y.~S.} \bibnamefont{Lee}},
  \bibinfo{journal}{Nature} \textbf{\bibinfo{volume}{492}},
  \bibinfo{pages}{406} (\bibinfo{year}{2012}), ISSN \bibinfo{issn}{0028-0836}.

\bibitem[{\citenamefont{Fu et~al.}(2015)\citenamefont{Fu, Imai, Han, and
  Lee}}]{Fu2015}
\bibinfo{author}{\bibfnamefont{M.}~\bibnamefont{Fu}},
  \bibinfo{author}{\bibfnamefont{T.}~\bibnamefont{Imai}},
  \bibinfo{author}{\bibfnamefont{T.-H.} \bibnamefont{Han}}, \bibnamefont{and}
  \bibinfo{author}{\bibfnamefont{Y.~S.} \bibnamefont{Lee}},
  \bibinfo{journal}{Science} \textbf{\bibinfo{volume}{350}},
  \bibinfo{pages}{655} (\bibinfo{year}{2015}).

\bibitem[{\citenamefont{Feng et~al.}(2017)\citenamefont{Feng, Li, Meng, Yi,
  Wei, Zhang, Wang, Jiang, Liu, Li et~al.}}]{Feng2017a}
\bibinfo{author}{\bibfnamefont{Z.}~\bibnamefont{Feng}},
  \bibinfo{author}{\bibfnamefont{Z.}~\bibnamefont{Li}},
  \bibinfo{author}{\bibfnamefont{X.}~\bibnamefont{Meng}},
  \bibinfo{author}{\bibfnamefont{W.}~\bibnamefont{Yi}},
  \bibinfo{author}{\bibfnamefont{Y.}~\bibnamefont{Wei}},
  \bibinfo{author}{\bibfnamefont{J.}~\bibnamefont{Zhang}},
  \bibinfo{author}{\bibfnamefont{Y.-C.} \bibnamefont{Wang}},
  \bibinfo{author}{\bibfnamefont{W.}~\bibnamefont{Jiang}},
  \bibinfo{author}{\bibfnamefont{Z.}~\bibnamefont{Liu}},
  \bibinfo{author}{\bibfnamefont{S.}~\bibnamefont{Li}}, \bibnamefont{et~al.},
  \bibinfo{journal}{Chinese Physics Letters} \textbf{\bibinfo{volume}{34}},
  \bibinfo{pages}{077502} (\bibinfo{year}{2017}),
  \urlprefix\url{http://stacks.iop.org/0256-307X/34/i=7/a=077502}.

\bibitem[{\citenamefont{Kelly et~al.}(2016)\citenamefont{Kelly, Gallagher, and
  McQueen}}]{Kelly2016}
\bibinfo{author}{\bibfnamefont{Z.~A.} \bibnamefont{Kelly}},
  \bibinfo{author}{\bibfnamefont{M.~J.} \bibnamefont{Gallagher}},
  \bibnamefont{and} \bibinfo{author}{\bibfnamefont{T.~M.}
  \bibnamefont{McQueen}}, \bibinfo{journal}{Phys. Rev. X}
  \textbf{\bibinfo{volume}{6}}, \bibinfo{pages}{041007} (\bibinfo{year}{2016}).

\bibitem[{\citenamefont{Guertler and Monien}(2011)}]{Guertler2011}
\bibinfo{author}{\bibfnamefont{S.}~\bibnamefont{Guertler}} \bibnamefont{and}
  \bibinfo{author}{\bibfnamefont{H.}~\bibnamefont{Monien}},
  \bibinfo{journal}{Phys. Rev. B} \textbf{\bibinfo{volume}{84}},
  \bibinfo{pages}{174409} (\bibinfo{year}{2011}),
  \urlprefix\url{https://link.aps.org/doi/10.1103/PhysRevB.84.174409}.

\bibitem[{\citenamefont{Guertler and Monien}(2013)}]{Guertler2013}
\bibinfo{author}{\bibfnamefont{S.}~\bibnamefont{Guertler}} \bibnamefont{and}
  \bibinfo{author}{\bibfnamefont{H.}~\bibnamefont{Monien}},
  \bibinfo{journal}{Phys. Rev. Lett.} \textbf{\bibinfo{volume}{111}},
  \bibinfo{pages}{097204} (\bibinfo{year}{2013}),
  \urlprefix\url{https://link.aps.org/doi/10.1103/PhysRevLett.111.097204}.

\bibitem[{\citenamefont{Jiang et~al.}(2017)\citenamefont{Jiang, Devereaux, and
  Kivelson}}]{Jiang2017}
\bibinfo{author}{\bibfnamefont{H.-C.} \bibnamefont{Jiang}},
  \bibinfo{author}{\bibfnamefont{T.}~\bibnamefont{Devereaux}},
  \bibnamefont{and} \bibinfo{author}{\bibfnamefont{S.~A.}
  \bibnamefont{Kivelson}}, \bibinfo{journal}{Phys. Rev. Lett.}
  \textbf{\bibinfo{volume}{119}}, \bibinfo{pages}{067002}
  (\bibinfo{year}{2017}),
  \urlprefix\url{https://link.aps.org/doi/10.1103/PhysRevLett.119.067002}.

\bibitem[{\citenamefont{Kresse and Hafner}(1993)}]{Kresse1993}
\bibinfo{author}{\bibfnamefont{G.}~\bibnamefont{Kresse}} \bibnamefont{and}
  \bibinfo{author}{\bibfnamefont{J.}~\bibnamefont{Hafner}},
  \bibinfo{journal}{Phys. Rev. B} \textbf{\bibinfo{volume}{47}},
  \bibinfo{pages}{558} (\bibinfo{year}{1993}),
  \urlprefix\url{https://link.aps.org/doi/10.1103/PhysRevB.47.558}.

\bibitem[{\citenamefont{Dronskowski and Bloechl}(1993)}]{Dronskowski1993}
\bibinfo{author}{\bibfnamefont{R.}~\bibnamefont{Dronskowski}} \bibnamefont{and}
  \bibinfo{author}{\bibfnamefont{P.~E.} \bibnamefont{Bloechl}},
  \bibinfo{journal}{The Journal of Physical Chemistry}
  \textbf{\bibinfo{volume}{97}}, \bibinfo{pages}{8617} (\bibinfo{year}{1993}),
  \eprint{http://dx.doi.org/10.1021/j100135a014},
  \urlprefix\url{http://dx.doi.org/10.1021/j100135a014}.

\end{thebibliography}
\end{document}